\documentclass[preprint,aps,pra,11pt
]{revtex4-1}
\usepackage{amssymb}
\usepackage{graphicx,color}
\usepackage{amsmath}
\usepackage{natbib}
\usepackage{epsfig}

\begin{document}

\title[Internal energy of one-component-plasma]{Internal energy of the classical two- and three-dimensional
one-component-plasma}

\author{Sergey A. Khrapak}
\affiliation{Aix-Marseille-Universit\'{e}, CNRS, Laboratoire PIIM, UMR 7345, 13397 Marseille cedex 20, France; \\
Forschungsgruppe Komplexe Plasmen, Deutsches Zentrum f\"{u}r Luft- und Raumfahrt,
Oberpfaffenhofen, Germany; \\
Joint Institute for High Temperatures, Russian Academy of Sciences, Moscow, Russia}

\author{Alexey G. Khrapak}
\affiliation{Joint Institute for High Temperatures, Russian Academy of Sciences, Moscow, Russia}

\begin{abstract}
We summarize several semi-phenomenological approaches to estimate the internal energy of one-component-plasma (OCP) in two (2D) and three (3D) dimensions. Particular attention is given to a hybrid approach, which reproduces
the Debye-H$\ddot{\text{u}}$ckel asymptote in the limit of weak coupling, the ion sphere (3D) and ion disc (2D) asymptotes in the limit of strong coupling, and provides reasonable interpolation between these two limits. More accurate ways to estimate the internal energy of 2D and 3D OCP are also discussed. The accuracy of these analytic results is quantified by comparison with existing data from numerical simulations. The relevance of the KTHNY theory in locating melting transition in 2D OCP is briefly discussed.
\end{abstract}

\maketitle                   

\section{Introduction}

The one-component-plasma (OCP) is an idealized system of identical point-like
charges  immersed in a uniform (rigid) neutralizing background of opposite charge\cite{Brush66,Baus80,Fortov06}. This model is of
considerable interest from the fundamental point of view and has wide interdisciplinary applications, including ionized matter in white dwarfs, interiors of heavy planets, alkali metals, colloidal suspensions, and complex (dusty) plasmas\cite{Fortov06,Brilliantov98,Fortov04,Fortov05,Ivlev12}. In addition, the OCP represents a very important example of classical systems of interacting particles with extremely soft interactions (the limit opposite to hard-sphere interactions) and as such
it plays significant role in condensed matter research.

In the three-dimensional (3D) case the interaction between the charged particles is described by the conventional Coulomb potential
\begin{equation}\label{Coulomb}
V(r)=Q^2/r,
\end{equation}
where $Q$ is the particle charge and $r$ is the distance between two particles. The system  is then characterized by the coupling parameter $\Gamma = Q^2/aT$, where
$T$ is the temperature (in energy units), $a = (4\pi n/3)^{-1/3}$ is the (3D) Wigner-Seitz radius, and $n$ is
the particle density. Thermodynamic properties of this system have been extensively studied in numerical simulations~\cite{Hansen73,Slattery80,Stringfellow90,Farouki94,Hamaguchi96,Dubin99,Caillol99,Caillol10}.

In the two-dimensional case (2D) two different systems are actually referred to as the OCP. The first is characterized by the conventional 3D Coulomb interaction potential (\ref{Coulomb}), but the particle motion is restricted to a 2D surface. This system has been used as a first approximation for the description of
electron layers bound to the surface of liquid dielectrics and of inversion layers in semi-conductor physics\cite{Baus80,Fortov06}. It has also some relevance to colloidal and complex (dusty) plasma mono-layers in the regime of week screening\cite{Fortov06,Fortov04,Fortov05,Ivlev12}. This system is characterized by the same coupling parameter as in 3D, except 2D Wigner-Seitz radius is used, $a=(\pi n)^{-1/2}$, where $n$ is now the 2D density. Thermodynamics of these systems has also been studied in the literature~\cite{Totsuji78,Gann79}.

There is another systems also referred to as the 2D OCP, in which the interaction potential is defined via the 2D Poisson equation and scales logarithmically with distance. The logarithmic potential, corresponding to the interaction of infinite charged filaments, is often employed to model interactions between vortices in thin-film superconductors. The 2D OCP has received considerable attention~\cite{Caillol82,Leeuw82,Choquard83,Radloff84} because of various field theoretical models\cite{Baus80} and existence of exact analytic solutions for some special cases~\cite{Jancovici81,Alastuey81}. The interaction potential between two particles follows from the solution of the 2D Poisson equation around a central test particle and reads
\begin{equation}
V(r)=-Q^2\ln(r/L),
\end{equation}
where $L$ is an arbitrary scaling length. It is common~\cite{Caillol82} to set $L=a$, where  $a$ is the 2D Wigner-Seitz radius. The thermodynamic of this system depends on the coupling parameter, $\Gamma=Q^2/T$, which is {\it density independent} (in contrast to Coulomb interactions in 3D and 2D).

The qualitative dependence of the OCP properties on the coupling strength is identical in 3D and 2D. As $\Gamma$ increases, the OCP shows a transition from a weakly coupled gaseous regime ($\Gamma\ll 1$) to a strongly coupled fluid regime ($\Gamma\gg 1$) and crystallizes at some $\Gamma_{\rm m}$ (the subscript ``m'' refers to melting). In the 3D case the stable crystalline phase is formed by the body-centered-cubic (bcc) lattice.
The transition occurs at $\Gamma_{\rm m}\simeq170 - 175$~\cite{Farouki94,Hamaguchi96,Dubin99,Khrapak14}. In the 2D case with the Coulomb interaction, numerical simulations located the transition into triangular lattice near $\Gamma_{\rm m}\simeq 125 \pm 15$~\cite{Gann79}. Experiments with a classical two-dimensional sheet of electrons yielded $\Gamma_{\rm m}\simeq 137 \pm 15$~\cite{Grimes79}. For the logarithmic interaction in 2D, numerical simulations and theory predict that the triangular lattice is thermodynamically favorable for $\Gamma\gtrsim 130 - 140$~\cite{Caillol82,Leeuw82,Choquard83,Radloff84}. The closeness of $\Gamma_{\rm m}$ values for these two different 2D systems is likely a coincidence, because the definitions of the coupling strength are different for logarithmic and Coulomb interactions. At even higher $\Gamma$, the glass transition has been predicted for 3D OCP~\cite{TanakaPRA_1987,YazdiPRE_2014}. This can also be a scenario for 2D OCP, but we are not aware of any work in this direction.

Thermodynamic properties of the OCP (in both 2D and 3D) have been extensively studied over decades and accurate numerical results as well as their fits are available in the literature.
Nevertheless, there has also been considerable continuous interest in deriving physically motivated analytical estimates or bounds on the thermodynamic quantities (in particular, internal energy) of the OCP. For example, analytical approaches of various complexity and accuracy have been discussed in Refs.~\cite{Brilliantov98,Caillol82,Leeuw82,Khrapak14,Mermin68,Gryaznov73,Lieb75,
Sari76,Stroud76,DeWitt79,Totsuji79,Deutsch79,Vieillefosse81,Rosenfeld82,Nordholm84,
Kaklyugin85,Ortner99,Caillol99_2}.
Below, we briefly remind some of the results particularly relevant to the present discussion.

Mermin \cite{Mermin68} demonstrated that the internal energy of the 3D OCP is bounded below by the Debye-H\"{u}ckel (DH) value. This demonstration is quite general and should be applicable to the 2D case with logarithmic interactions, too. This bound is a reasonable measure of the actual OCP energy at weak coupling. Lieb and Narnhofer \cite{Lieb75} derived another exact lower bound on the reduced energy (energy per particle in units of system temperature) of the 3D OCP, which reads $u> -0.9\Gamma$. This result is often refereed to as the ion sphere model \cite{Baus80,Ichimaru82} (ISM) and provides rather good estimate of the internal energy at strong coupling. Similar lower bound has been identified for the 2D OCP with logarithmic interactions by Sari and Merlini~\cite{Sari76}. It reads $u> -0.375 \Gamma$ and is usually referred to as the ion disc model (IDM). Again, IDM is surprisingly accurate at strong coupling. Gryaznov and Iosilevskiy \cite{Gryaznov73}, and later independently Nordholm \cite{Nordholm84}, proposed a simple modification of the DH theory for 3D, called "DH plus hole" (DHH) approximation, based on the recognition that the exponential particle density must be truncated close to the particles so as not to become negative. It improves considerably the DH theory at moderate coupling, $\Gamma\lesssim 1$, but exhibits improper scaling ($\propto -0.75\Gamma$) in the high-$\Gamma$ limit. This approach can be extended to the 2D OCP with logarithmic interactions, as we demonstrate below. More recently, Caillol derived two other exact lower bounds for the internal energy\cite{Caillol99_2}, which have been demonstrated to be in better agreement with the numerical results than those obtained previously in a wide range of $\Gamma$.

The purpose of this paper is threefold. First, we summarize yet another simple analytical scheme to estimate the internal energy of the OCP in 3D and 2D. The approach is based on
the hybrid DHH + ISM/IDM consideration formulated below. Simple electrostatic consideration, involving the solution of the Poisson equation, is used and thus, in 2D case, the approach  is limited to the logarithmic interaction. It produces expressions, which reduce to the DH result at weak coupling and to the ISM/IDM results at strong coupling and provide reasonable interpolation between these limits. Second, we briefly summarize simple and accurate fits for all three OCP systems discussed here. In particular, we demonstrate that in the 2D OCP the thermal component of the internal energy exhibits the same scaling for Coulomb and logarithmic interactions. Similar scaling also holds for Yukawa interactions near the OCP limit (long screening length) and this suggests that it is a universal property of soft repulsive particle systems in 2D. Based on these accurate scalings of the internal energy other thermodynamic properties can be easily calculated. Finally, we briefly compare the location of the fluid-crystal phase transition in 2D OCP with Coulomb and logarithmic interactions, as estimated using the Kosterlitz-Thouless-Halperin-Nelson-Young (KTHNY) theory.

\section{Hybrid approach to the internal energy of the OCP in 3D and 2D}

\subsection{Linear Debye-H\"{u}ckel approximation.}

The solution of the linear Poisson-Boltzmann equation, $\Delta\phi = k_{\rm D}^2\phi$, in 3D and 2D yields
\begin{equation}\label{phia}
 \phi(r)=
\left\{\begin{aligned}
\frac{Q}{r}e^{-k_{\rm D}r},\qquad k_{\rm D}=\sqrt{4\pi n Q^2/T},\qquad (3\rm D) \\
Q K_0(rk_{\rm D}),\qquad k_{\rm D}=\sqrt{2\pi n Q^2/T},\qquad (2\rm D)
\end{aligned} \right.
\end{equation}
where $K_0(x)$ is the zero-order modified Bessel function of the second kind and $k_{\rm D}$
is the inverse screening length. Note the relations $k_{\rm D}a=\sqrt{3\Gamma}$ in 3D and $k_{\rm D}a=\sqrt{2\Gamma}$ in 2D. The reduced excess (that over non-charged particles) energy of the systems, independently of dimensionality, can be evaluated from
\begin{equation}\label{energy}
u_{\rm ex}\equiv \frac{U_{\rm ex}}{NT}=\frac{\left[Q\phi(r)-V(r)\right]_{r\rightarrow 0}}{2T},
\end{equation}
where $N$ is the number of the particles ($N\rightarrow \infty$ in the thermodynamic limit). This corresponds to the DH approximation for the weakly coupled ($\Gamma\ll 1$) limit:
\begin{equation}\label{uDH}
u_{\rm DH}(\Gamma)=
\left\{\begin{aligned}
-\frac{\sqrt{3}}{2}\Gamma^{3/2},\qquad (3\rm D) \\
-\frac{\Gamma}{4}\left(\ln \frac{\Gamma}{2}+2\gamma\right),\qquad (2\rm D)
\end{aligned} \right.
\end{equation}
where $\gamma\simeq 0.57721$ is the Euler's constant (we used the expansion $K_0(x)\simeq -\gamma +\ln2 -\ln x +{\mathcal O}(x^2)$ for $x\ll 1$). The DH approximation provides accurate results only in the limit of extremely weak coupling.

\subsection{Debye-H\"{u}ckel plus hole approximation}

To extend the applicability of the DH approach to the moderately coupled OCP, the simple phenomenological ``Debye-H\"{u}ckel plus hole'' (DHH) approximation was proposed~\cite{Nordholm84,Gryaznov73}. The main idea behind the DHH approximation is that the exponential particle density must be truncated close to a test particle in order to avoid density to be negative upon linearization. The DHH approach was originally applied to the 3D OCP. Here we outline its application to the 2D case, but 3D results are also summarized for completeness.

\begin{figure}[t]
 \begin{center}
\includegraphics [width=7cm]{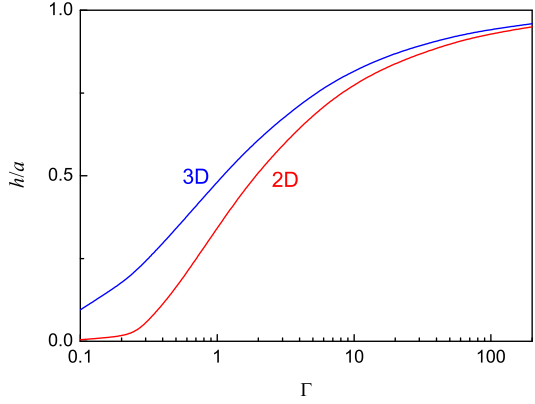}
\end{center}
\caption{Reduced radius of the hole, $h/a$, around the test particle as a function of the coupling parameter $\Gamma$ in the 3D and 2D OCP.}
\label{h1}
\end{figure}

The potential inside the hole (sphere in 3D and disk in 2D cases) of radius $h$ can be written as
\begin{equation}\label{in}
\phi_{\rm in}(r)=
\left\{\begin{aligned}
\frac{Q}{r}+{\mathcal A}_0+{\mathcal A}_2r^2,\qquad (3\rm D) \\
-Q\ln(r/a)+{\mathcal A}_0+{\mathcal A}_2r^2.\qquad (2\rm D)
\end{aligned} \right.
\end{equation}
Outside the hole, the potential satisfies the linearized Poisson-Boltzmann equation, so that
\begin{equation}\label{out}
\phi_{\rm out}(r)=
\left\{\begin{aligned}
\frac{{\mathcal B}}{r}e^{-k_{\rm D}r},\qquad (3\rm D) \\
{\mathcal B}K_0(rk_{\rm D}).\qquad (2\rm D)
\end{aligned} \right.
\end{equation}
The two solutions should be matched at $r=h$, requiring $ \phi_{\rm in}(h)=\phi_{\rm out}(h)=T/Q$ (the last condition ensures that particle density vanishes at the hole boundary in the linear approximation) and $\phi'_{\rm in}(h)=\phi'_{\rm out}(h)$. Using the identity $K_0'(x)=-K_1(x)$ we get the following  equations for $z=h/a$
\begin{equation}\label{za}
z^2=
\left\{\begin{aligned}
\frac{1}{3\Gamma}\left\{\left[1+(3\Gamma)^{3/2}\right]^{1/3}-1\right\}^2,\qquad (3\rm D) \\
1-z\sqrt{\frac{2}{\Gamma}}\frac{K_1(\sqrt{2\Gamma}z)}{K_0(\sqrt{2\Gamma}z)}.\qquad (2\rm D)
\end{aligned} \right.
\end{equation}
 Unlike the 3D case, where the hole radius is expressed explicitly in terms of $\Gamma$, in the 2D case transcendent equation should be solved numerically. This, however, does not represent a major difficulty. The dependence $z(\Gamma)$ is shown in Fig.~\ref{h1} for both 3D and 2D OCP. In both cases $h\rightarrow 0$ when $\Gamma\rightarrow 0$, and $h\rightarrow a$ when $\Gamma\rightarrow\infty$.

\begin{figure}
\begin{minipage}{85mm}
\includegraphics[width=6.9cm]{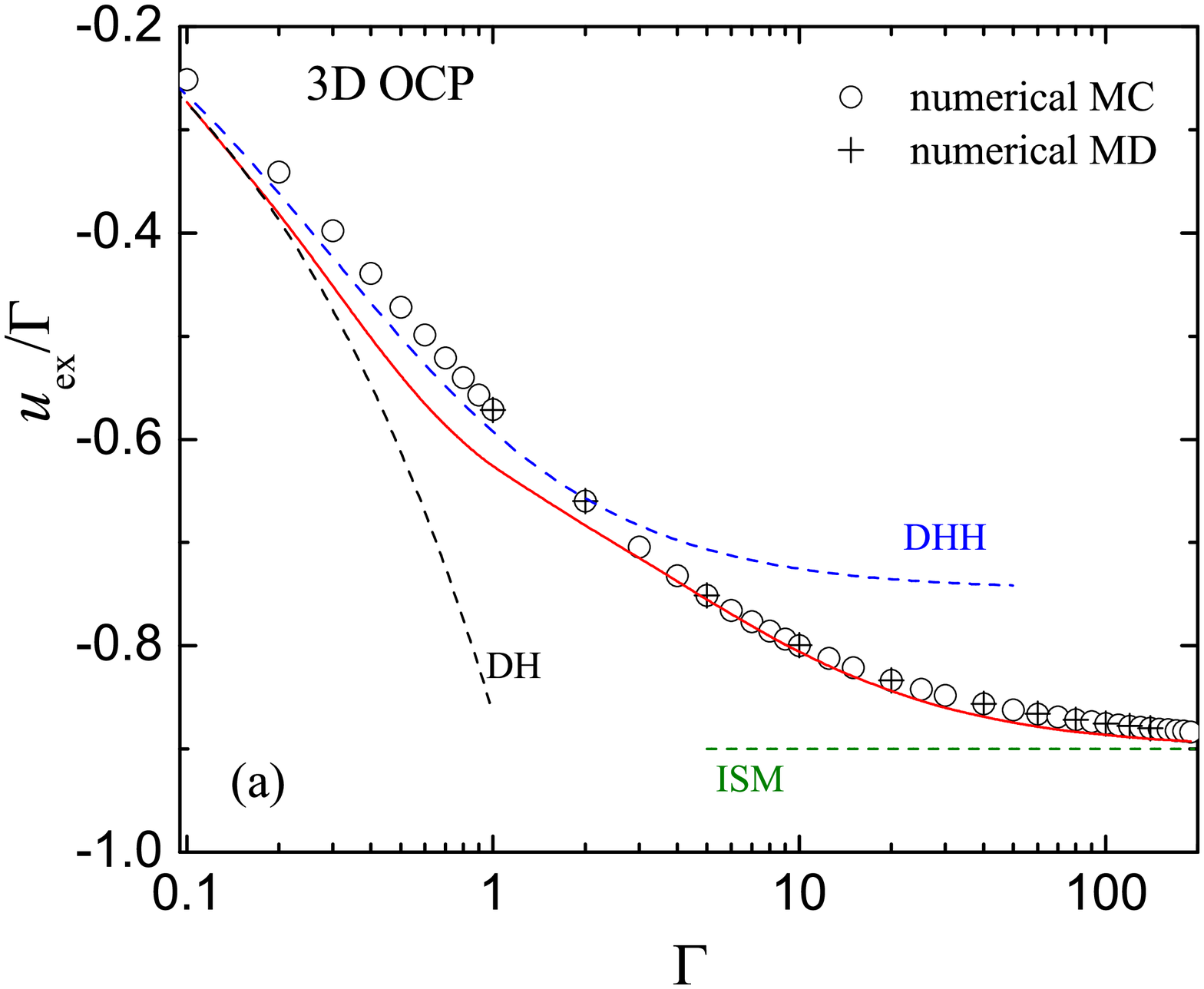}
\end{minipage}
\hfil
\begin{minipage}{85mm}
\includegraphics[width=7.cm]{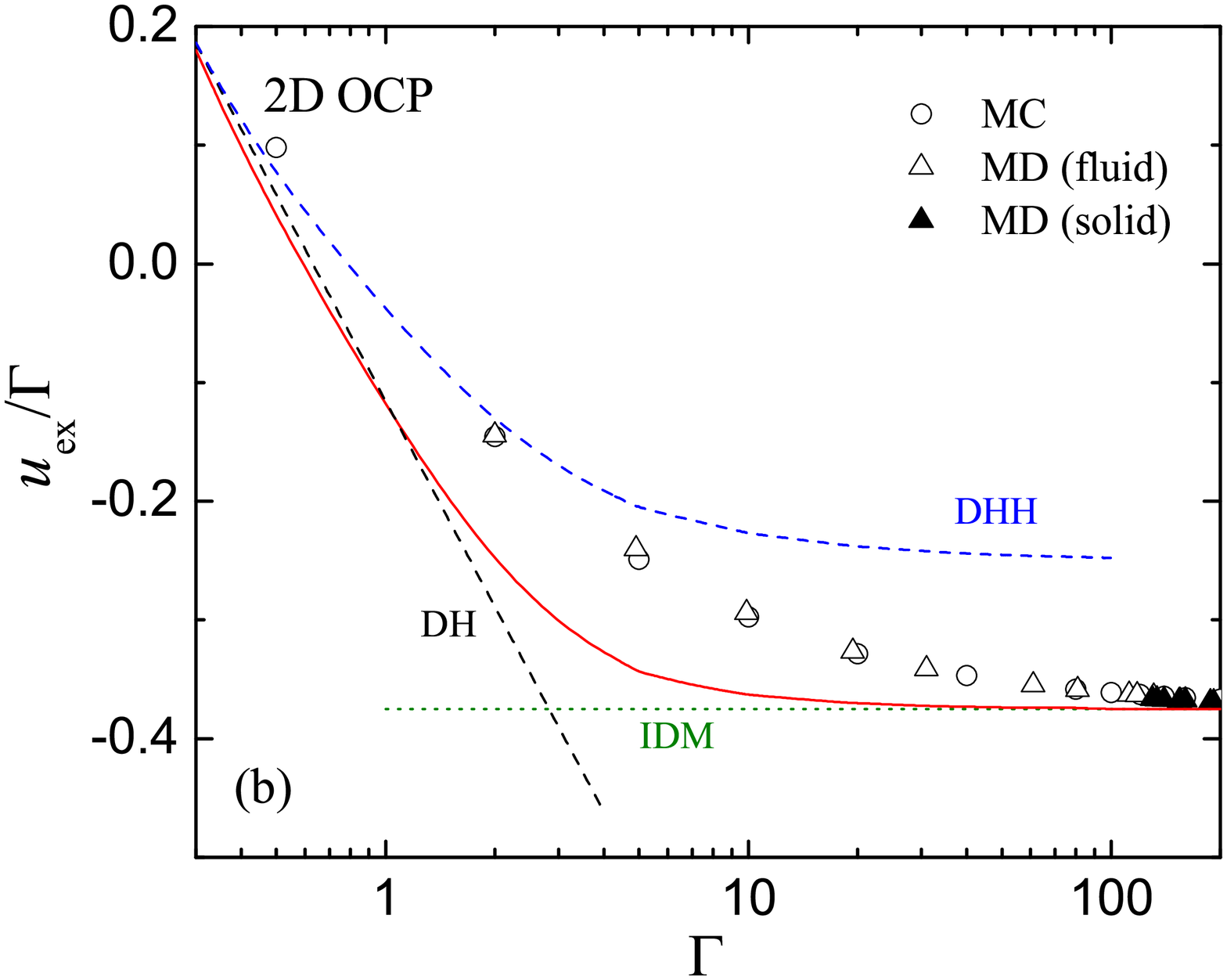}
\end{minipage}
\caption{Reduced excess energy $u_{\rm ex}/\Gamma$ versus the coupling parameter $\Gamma$ for the 3D OCP (a) and 2D OCP (b). For the 3D case, symbols are the results from numerical MC~\cite{Caillol99,Caillol10} and MD~\cite{Farouki94,Hamaguchi96} simulations. Similarly, for the 2D case, symbols are the results from MC~\cite{Caillol82} and MD~\cite{Leeuw82} simulations. Various  curves correspond to the DH, DHH, ISM and IDM approximations, as indicated in the figures. The (red) solid curves in both figures show the result of the hybrid DHH+ISM (3D) and DHH+IDM (2D) approximation of Eq.~(\ref{Hybrid}).}
\label{u2}
\end{figure}

The reduced excess energy can be evaluated using equation (\ref{energy}), which yields $u_{\rm DHH}=(Q{\mathcal A}_0/2T)$. This results in
\begin{equation}\label{uexb}
u_{\rm DHH}(\Gamma)=
\left\{\begin{aligned}
-\frac{1}{4}\left\{\left[1+(3\Gamma)^{3/2}\right]^{2/3}-1\right\},\qquad (3\rm D) \\
\frac{1}{2}+\frac{\Gamma}{2}\ln z-\frac{\Gamma}{4}z^2. \qquad (2\rm D)
\end{aligned} \right.
\end{equation}
In the limit $\Gamma\ll 1$, Eq.~(\ref {uexb}) reduces to the DH results of Eq.~(\ref {uDH}), but it remains adequate at higher $\Gamma$ than the DH approach does. For example, in the 2D OCP the exact result can be obtained analytically in the special case $\Gamma=2$~\cite{Jancovici81,Alastuey81}. The exact excess energy at this point is $u_{\rm ex}(2)=-\gamma/2\simeq -0.28861$~\cite{Jancovici81}. The DHH value is very close to that, $u_{\rm DHH}(2)\simeq -0.29324$, while the DH value is considerably below the exact one, $u_{\rm DH}(2)\simeq - 0.57721$.  In the strongly coupled regime $\Gamma\gg1$, the DHH approximation yields the correct scaling $u_{\rm ex} \propto \Gamma$, but the coefficient of proportionality is incorrect ($-0.75$ instead of $\simeq -0.9$ in 3D and $-0.25$ instead of $\simeq -0.375$ in 2D). In Figure~\ref{u2} we compare the energies obtained using the DHH approach with those obtained using Monte Carlo (MC) and molecular dynamics (MD) computer simulations. It is worth noting that the application of the DHH approach to 3D Yukawa systems has been recently discussed in Ref.~\cite{DHH} in the context of complex (dusty) plasmas.

\subsection{Ion sphere and ion disc models}

 The main idea of the ion sphere (ISM) and ion disk (IDM) models is that in the regime of strong coupling, the particles repel each other and form a regular structure with the interparticle spacing of order $a$\cite{Baus80}. Each particle can be considered as restricted to the cell (sphere in 3D and disc in 2D) of radius $a$, filled with the neutralizing background. The cells are charged neutral and do not overlap, and hence the potential energy of the system is just the sum of potential energy of each cell. The latter is readily calculated from the pure electrostatic consideration~\cite{Lieb75,Sari76}. The result is
 \begin{equation}\label{ISDM}
u_{\rm ISM/IDM}=
\left\{\begin{aligned}
-\frac{9}{10}\Gamma= -0.9\Gamma,\qquad (3\rm D) \\
-\frac{3}{8}\Gamma= -0.375\Gamma.\qquad (2\rm D)
\end{aligned} \right.
\end{equation}
The results are very close to the static components of the actual excess energy of the 3D and 2D OCP in both strongly coupled fluid and solid phases. They can be compared with
the Madelung constants of the OCP bcc lattice, $u_{\rm M}=-0.895929\Gamma$ (3D), and triangular lattice, $u_{\rm M}=-0.37438\Gamma$ (2D). The agreement is impressive. It was proven mathematically that Eqs.~(\ref{ISDM}) provides the lower bounds of the excess internal energy in the thermodynamic limit \cite{Lieb75,Sari76}. The ISM and IDM asymptotes are shown in Fig.~\ref{u2}. The ISM model can be easily generalized to 3D Yukawa systems. It is worth to mention that the ISM result for the excess energy can also be obtained from the energy equation using Percus-Yevick (PY) radial distribution function for hard spheres at the unphysical packing fraction $\eta=1$, which provides some link between the ISM approximation and the integral equation theories (for details see Ref.~\cite{ISM} and references therein).

\subsection{Hybrid approximation}

Now we discuss the recently proposed hybrid approach to the excess energy of 3D and 2D OCP, which tends to reproduce the DH and ISM (IDM) results in the respective limits of weak and strong coupling, and provides reasonable interpolation between these limits~\cite{Hybrid3D,Hybrid2D}.

Let us consider a test particle along with the piece of the neutralizing background charge (sphere or disc of radius $h$ in 3D or 2D, correspondingly) as a new compound particle. The internal energy of such a compound particle consists of two parts: energy of a uniformly charged cell of radius $h$ and charge $q=-Q(h/a)^D$ ($D$ is the system dimension) and the energy of a charge $Q$ placed in the center of such a cell. Solving the Poisson equation inside and outside the cell and matching the solutions we get for the energy of the uniformly charged cell of background charge
 \begin{equation}\label{ub}
u_{\rm b}=
\left\{\begin{aligned}
\frac{3}{5}\frac{q^2}{Th},\qquad (3\rm D) \\
\frac{q^2}{T}\left(\frac{1}{8}-\frac{1}{2}\ln\frac{h}{a}\right). \qquad (2\rm D)
\end{aligned} \right.
\end{equation}
The energy of a charge $Q$ placed in the center of such a cell is
 \begin{equation}\label{up}
u_{\rm p}=
\left\{\begin{aligned}
\frac{3}{2}\frac{qQ}{Th},\qquad (3\rm D) \\
\frac{qQ}{T}\left(\frac{1}{2}-\ln\frac{h}{a}\right). \qquad (2\rm D)
\end{aligned} \right.
\end{equation}
The energy of the compound particle is then
 \begin{equation}\label{ucp}
u_{\rm cp}(\Gamma)=
\left\{\begin{aligned}
\Gamma z^2\left(\frac{3}{5}z^3-\frac{3}{2}\right),\qquad (3\rm D) \\
\Gamma z^2\left(\ln z-\frac{1}{2}\right)+\Gamma z^4\left(\frac{1}{8}-\frac{1}{2}\ln z\right).\qquad (2\rm D)
\end{aligned} \right.
\end{equation}
In the limit of strong coupling, the effective charge of the compound particle tends to zero and, therefore, its internal energy should be an adequate measure of the excess energy of the whole system (per particle). We get in this limit $z\rightarrow 1$ and  $u_{\rm cp}\simeq-0.9\Gamma$ (3D) or $u_{\rm cp}\simeq-0.375\Gamma$ (2D), which coincides with the ISM/IDM results.

The energy associated with the remaining interaction between the compound particles (they are not charge neutral in the general case) can be estimated from the energy equation
\begin{equation}\label{uppa}
u_{\rm pp}=(n/2T)\int_{r>h} V_{\rm eff}(r)[g(r)-1]d{\bf r},
\end{equation}
where $V_{\rm eff}(r)$ is the Coulomb (3D) or logarithmic (2D) interaction potential between the compound particles with effective charge $Q_{\rm eff}=Q+q=Q[1-z^D]$ and $g(r)$ is the radial distribution function. Since the effective charge $Q_{\rm eff}$  is considerably reduced compared to the actual charge $Q$, especially in the strong coupling regime, it is not very unreasonable to use an expression originating from the linearized Boltzmann relation, $g(r) \simeq 1-Q_{\rm eff}\phi_{\rm out}(r)/T$, where $\phi_{\rm out}$ is given by Eq. (\ref{out}) in the DHH approximation. This yields
\begin{equation}\label{uppb}
u_{\rm pp}(\Gamma)=
\left\{\begin{aligned}
-\frac{\sqrt{3\Gamma}}{2}z(1-z^3)^3,\; \qquad (3\rm D) \\
\frac{\Gamma(1-z^2)^3}{K_0(\sqrt{2\Gamma}z)}\int_{z}^{\infty}x\ln x K_0(\sqrt{2\Gamma}x)dx.\; \qquad (2\rm D)
\end{aligned} \right.
\end{equation}
In the 3D case, we have at weak coupling $z\sim \Gamma$ and the DH result is immediately recovered. In the 2D case numerical integration is generally required in (\ref{uppb}), but it can be also shown that the result reduces to the DH one in the weakly coupled limit ($\Gamma\ll 1$).

Our estimate for the OCP excess energy within the hybrid DHH+ISM/IDM approximation is then simply
\begin{equation}\label{Hybrid}
u_{\rm hyb}(\Gamma)=u_{\rm cp}(\Gamma)+u_{\rm pp}(\Gamma).
\end{equation}
Equation~(\ref{Hybrid}) reduces to the DH and ISM/IDM asymptotes in respective limits of weak and strong coupling. The quality of the interpolation between these two limits is illustrated in Fig.~\ref{u2} (red solid curves). The agreement with the accurate numerical data from MC and MD simulations is better in the 3D case, but remains also acceptable in the 2D case, taking into account the simplicity of the model. However, it is also obvious that in many situations this accuracy is insufficient, and we summarize more accurate expressions in the next Section.

\section{Accurate expressions for the internal energy of OCP fluids}

For strongly coupled systems, the reduced excess energy  can be conveniently divided into static and thermal components
\begin{equation}\label{uex}
u_{\rm ex}=u_{\rm st}+u_{\rm th}.
\end{equation}
The static contribution corresponds to the value of internal energy when the particles are frozen in some regular configuration (e.g., crystalline lattice for solids), and the thermal corrections arise due the deviations from these fixed positions, associated with thermal fluctuations. When the value of the static component of the excess energy is specified, the thermal component determines the excess energy and other thermodynamic properties of the system. It is known that the thermal energy exhibits quasi-universal scaling for soft repulsive interactions in 3D (first proposed by Rosenfeld and Tarazona (RT)~\cite{Rosenfeld98,Rosenfeld00}) and two interesting questions arise: (i) How accurate is the RT scaling for 3D OCP; and (ii) Whether there is some analog of the RT scaling in 2D.

\begin{figure}[t]
\begin{center}
\includegraphics [width=8.3cm]{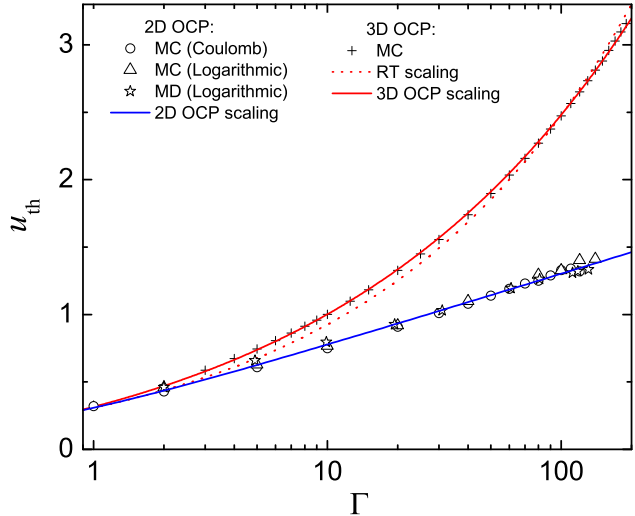}
\end{center}
\caption{Thermal component of the reduced excess energy, $u_{\rm th}$, of the strongly coupled 3D and 2D OCP fluids versus the coupling parameter $\Gamma$. Symbols correspond to MC and MD simulations: Crosses are MC results for 3D OCP~\cite{Caillol99}, circles are MC results for 2D OCP with the Coulomb interaction~\cite{Gann79}, triangles and stars are MC and MD results for 2D OCP with logarithmic interaction, respectively~\cite{Caillol82,Leeuw82}. The red solid curve is the 3D OCP fit of Eq.(\ref{uth}).
The blue solid line corresponds to the 2D OCP scaling. The red dashed curve represents the RT scaling of Eq.(\ref{uRT}).}
\label{fig3}
\end{figure}

First, let us consider the 3D case. A relevant measure of the static energy component in 3D OCP fluids is the ISM energy (we remind that it is very close to the bcc lattice sum).
Based on the accurate MC simulation results from Ref.\cite{Caillol99}, a simple two-term expression for $u_{\rm th}$ has been proposed~\cite{Khrapak14}
\begin{equation}\label{uth}
u_{\rm th}=0.5944\Gamma^{1/3}-0.2786.
\end{equation}
This fit, along with the MC numerical data, is plotted in Fig.~\ref{fig3}. In addition, a variant of the RT scaling,
\begin{equation}\label{uRT}
u_{\rm th}\simeq3.2(\Gamma/\Gamma_{\rm m})^{2/5}-0.1,
\end{equation}
is also plotted. This scaling has been successfully used to obtain practical expressions for the internal energy and pressure of 3D Yukawa fluids~\cite{KhrapakPRE15,KhrapakJCP15,KhrapakPPCF15} in a relatively wide range of screening strength. These practical expressions can be for instance used to estimate the sound velocity of Yukawa fluids with applications to complex (dusty) plasmas~\cite{KhrapakPRE_DA}. Figure \ref{fig3} demonstrates that the RT scaling describes fairly well the numerical data, but the OCP scaling is somewhat more accurate, especially in the regime $1\leq\Gamma\leq100$. This is absolutely not surprising, since the OCP scaling is nothing but the best simple fit to the numerical data shown in Fig. \ref{fig3}.  The very fact that the exponent $s=1/3$  (or close to that) in Eq.~(\ref{uth}) provides
particularly good agreement with the numerical data for the OCP has been documented in a number of previous studies~\cite{Farouki94,Hamaguchi96,Caillol99,Stringfellow90,Dubin99}. It is worth noting that the OCP expression (\ref{uth}), rewritten in the RT-like form (i.e., $u_{\rm th}$ expressed in terms of $\Gamma/\Gamma_{\rm m}$) is also superior to RT scaling for Yukawa fluids near the OCP limit (when screening length is longer or comparable to the inter-particle spacing). This provides us with a simple and accurate practical tool to estimate thermodynamic properties of weakly screened Yukawa fluids~\cite{KhrapakPoP15}. 

Let us now consider the 2D case. The energy can be again divided into the static and thermal parts, according to Eq. (\ref{uex}).    
For the static component of the energy in the 2D case, we now chose the triangular lattice sums (Madelung energies), which are $u_{\rm M}=-0.37438\Gamma$ for the logarithmic and $u_{\rm M}=-1.106103\Gamma$  for the Coulomb potential, respectively \cite{Caillol82,Gann79} (the ion disc model can be constructed for logarithmic interactions as we discussed above, but we are not aware of any such construction for the Coulomb interaction in 2D). Then, subtracting the static component from the full excess energy, available from the previous numerical simulations, we can obtain the thermal energy component. The resulting dependence of $u_{\rm th}$ on the coupling parameter $\Gamma$ is shown in Fig.~\ref{fig3}. The numerical data points, for both Coulomb and logarithmic interactions, tend to collapse on a single quasi-universal curve. 
Some scattering of the data points is present, but no clear systematic trend is observed, indicating that this may simply reflect the level of accuracy of the simulation results (note that at large $\Gamma$, the thermal component is a tiny fraction of the total OCP excess energy). The dependence $u_{\rm th} (\Gamma)$ has a logarithmic character, the blue solid line corresponds to 
\begin{equation}\label{uth_2D}
u_{\rm th}\simeq 0.231\ln(1+2.798\Gamma).
\end{equation}
Since the values of $\Gamma_{\rm m}$ for the 2D OCP with Coulomb and logarithmic interactions are rather close, an analog of the RT scaling for soft repulsive particles in 2D emerges. The expression proposed in Ref.~\cite{KhrapakPoP15} is
\begin{equation}\label{utha}
 u_{\rm th}\simeq b_1\ln[1+b_2(\Gamma/\Gamma_{\rm m})],
\end{equation}     
 with the coefficients $b_1=0.231$ and $b_2 =391.655$. Similarly to the 3D case, this scaling is well applicable to weakly screened Yukawa systems in 2D. Various thermodynamic functions of 2D Yukawa fluids can be easily estimated. Good agreement with the simulation results of Refs.~\cite{Totsuji04,Hartmann05} has been documented~\cite{KhrapakPoP15}. Applications to estimate the sound velocity in two-dimensional Yukawa fluids have been discussed in Ref.~\cite{2Dsound}

\section{Free energy and pressure}

Having relatively accurate expressions for the excess internal energy of strongly coupled OCP in 3D and 2D, we can evaluate other thermodynamic quantities. Here we present expressions for the reduced Helmholtz free energy and pressure. The generic expression for the reduced excess (that over non-interacting particles) free energy of the strongly coupled OCP fluid is
\begin{equation}\label{free}
f_{\rm fluid}(\Gamma)=f(\Gamma_0)+\int_{\Gamma_0}^{\Gamma}d\Gamma' u_{\rm ex}(\Gamma')/\Gamma',
\end{equation}
where $\Gamma_0$ corresponds to the weakly or moderately coupled regime, and $f(\Gamma_0)$ is known to a good accuracy. In some cases $\Gamma_0$ can be simply set zero, because the exact behavior of the excess energy at weak coupling normally has very little effect on the excess free energy at strong coupling. This is, however, not the case for the 3D OCP, because the integral in Eq.~(\ref{free}) diverges when expression (\ref{uth}) is used. Therefore, we chose $\Gamma_0=1$, $f(\Gamma_0)=-0.4368$~\cite{Farouki94} to get
\begin{equation}
f_{\rm fluid}=-0.9\Gamma+1.7832\Gamma^{1/3}- 0.2786\ln\Gamma-1.3200.
\end{equation}
This expression has been previously derived in Ref.~\cite{Khrapak14}. For the 2D OCP the scaling (\ref{uth_2D}) does not lead to the integral divergence at small $\Gamma$. Although it also does not reproduce the exact behavior of $u_{\rm ex}$ in the limit of weak coupling, for strongly coupled 2D OCP with Coulomb interactions, we find appropriate to put $\Gamma_0=0$, which yields
\begin{equation}
f_{\rm fluid}=-1.106103\Gamma -0.231{\rm Li}_2(-2.798\Gamma),
\end{equation}
where ${\rm Li}_2(z)=\int_z^0dt \ln(1-t)/t$ is dilogarithm. For the 2D OCP with logarithmic interaction we should use the fact that the reduced free energy at $\Gamma_0=2$ is known exactly, $f(2)= 1-\tfrac{1}{2}\ln(2\pi)\simeq 0.0811$~\cite{Alastuey81}. We have then
\begin{equation}
f_{\rm fluid}=-0.37438\Gamma -0.231{\rm Li}_2(-2.798\Gamma)+0.1469.
\end{equation}

To check the accuracy of these expressions we look for the intersection of the fluid and solid free energies of the considered systems. This is a very stringent test, since the free energies of fluid and solid are nearly parallel near the intersection. For the free energy of the 3D OCP forming the bcc lattice we take the expression from Ref.~\cite{Dubin99},
\begin{equation}
f_{\rm solid}=-0.895929\Gamma+\frac{3}{2}\ln\Gamma-1.1704-\frac{10.84}{\Gamma}-\frac{352.8}{2\Gamma^2}-\frac{179400}{3\Gamma^3},
\end{equation}
where the last three terms represent anharmonic corrections.
For the 2D OCP forming the triangular lattice we use the result from Ref.~\cite{Gann79}
\begin{equation}
f_{\rm solid}=-1.106103\Gamma+\ln\Gamma-\ln2+0.298-\frac{5}{\Gamma}-\frac{560}{2\Gamma^2},
\end{equation} 
where the last two terms are again anharmonic corrections. Finally, for the 2D OCP solid with the logarithmic interaction we use the available result of a simple harmonic approximation~\cite{Alastuey81},
\begin{equation}
f_{\rm solid}=-0.37438\Gamma+\ln\Gamma-0.262.
\end{equation}

\begin{figure}[t]
\begin{center}
\includegraphics [width=15.cm]{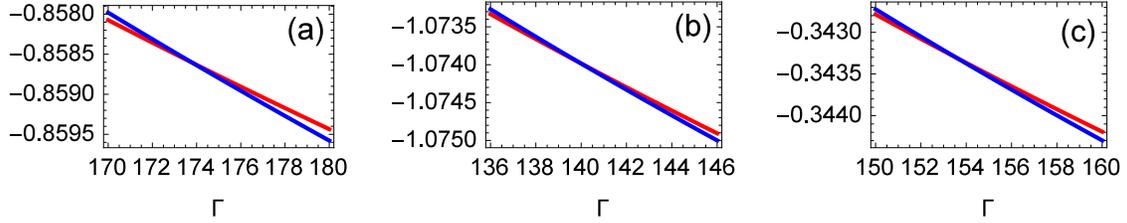}
\end{center}
\caption{Reduced excess free energy, in units of $\Gamma$ (i. e. $f_{\rm ex}/\Gamma$), as a function of the coupling parameter $\Gamma$ of 3D OCP (a), 2D OCP with Coulomb interaction (b), and 2D OCP with logarithmic interaction (c). The red curves correspond to the fluid phase, the blue curves correspond to the crystalline solid. Their intersection locates the point of the fluid-crystal phase transition. This figure gives yet another illustration that the free energies of the fluid and solid phases are nearly parallel in the vicinity of their intersection, indicating that very high accuracy is required to properly determine the location of the phase transition.}
\label{fig5}
\end{figure}

The fluid and solid reduced excess free energies near their intersection are shown in Fig.~\ref{fig5}. From the location of the intersection point we can estimate the coupling parameter at the fluid-crystal phase transition. For the 3D OCP we get $\Gamma_{\rm m}\simeq 174$ in very good agreement with the result of Ref.~\cite{Dubin99}. For the 2D OCP with the Coulomb interaction we get $\Gamma_{\rm m}\simeq 140$, which is consistent with the range predicted in earlier numerical simulations and experiments. For the 2D OCP with the logarithmic interaction, intersection occurs near $\Gamma_{\rm m}\simeq 154$, which is somewhat higher than obtained in Refs.~\cite{Caillol82,Leeuw82}. However, we remind that the free energy of the solid phase has been evaluated using a simple harmonic approximation. By analogy with the two other OCP systems, it can be expected that if the anharmonic corrections are properly accounted for, the coupling parameter corresponding to the phase transition can decrease to $\Gamma_{\rm m}\simeq 130 - 140$, in much better agreement with the results of previous studies. 

Regarding the excess pressure, it can be trivially obtained using the virial (pressure) equation. For the OCP with Coulomb interactions we have, in reduced units, $p_{\rm ex}=\tfrac{1}{3}u_{\rm ex}$ in 3D and $p_{\rm ex}=\tfrac{1}{2}u_{\rm ex}$ in 2D. For the 2D OCP with the logarithmic potential, the virial equation combined with the charge neutrality condition immediately yields $p_{\rm ex}=-\tfrac{1}{4}\Gamma$. This simple exact result is a consequence of the observation that density is an irrelevant variable for 2D OCP with the logarithmic potential~\cite{Caillol82}. 

Other thermodynamics quantities can be evaluated in a similar way.     

\section{Melting of the 2D OCP in the KTHNY model}

The Kosterlitz-Thouless-Halperin-Nelson-Young
(KTHNY) theory~\cite{KTHNY} describes melting in classical 2D systems with arbitrary interaction between the particles. Recently, the KTHNY scenario has been confirmed for sufficiently soft interactions~\cite{Kapfer15}, indicating that it should be relevant to 2D OCP, both with Coulomb and logarithmic interactions. The KTHNY theory states that the melting transition occurs when
\begin{equation}
\frac{4 \pi T}{b^2}= \frac{\mu(\mu+\lambda)}{2\mu+\lambda},
\end{equation}
where $b$ is the lattice spacing and $\mu$, $\lambda$ are the Lam\'{e} coefficients of 2D solid, which can be expressed via the longitudinal and transverse sound velocities~\cite{Peeters87}. For the OCP the longitudinal sound velocity is infinite and the melting temperature can be expressed via the transverse sound velocity~\cite{Thouless78}
\begin{equation}
T_{\rm m}=\frac{mnb^2c_{\rm T}^2}{4\pi},
\end{equation} 
where $mn$ is the mass per unit area and $c_{\rm T}$ is the transverse sound velocity.
For the triangular lattice we have $b=(2/n\sqrt{3})^{1/2}$. Then, for the Coulomb interaction, using the zero-temperature limit $c_{\rm T}\simeq 0.513\sqrt{Q^2/mb}$~\cite{Peeters87} we get $\Gamma_{\rm m}\simeq 79$, as obtained originally by Thouless~\cite{Thouless78}. This is clearly only a rough estimate of the melting location, considerable improvement can be achieved by taking into account the temperature dependence of the shear modulus arising from the phonon-phonon interaction and the polarizability of dislocation pairs~\cite{Morf79}. Nevertheless, it is interesting to make similar estimation for the 2D OCP with the logarithmic interaction. We use the transverse sound velocity $c_{\rm T}=\sqrt{Q^2/8m}$ derived in Ref.~\cite{Alastuey81} to get
\begin{equation}
\Gamma_{\rm m}=16\pi\sqrt{3}\simeq 87.
\end{equation}               
This estimate demonstrates the same level of accuracy as in the Coulomb case, improvements seem necessary.  

\section{Conclusion}

The one-component-plasma is an old model with wide iterdisciplinary applications. It also represents an important example of classical systems with extremely soft interactions. In this paper we mainly discussed thermodynamic properties of the model (in terms of the internal energy). Of particular significance can be the observation that the OCP scaling of the thermal component of the excess energy exhibits scaling, which is quasi-universal and applies to other soft repulsive potentials, both in 2D and 3D cases. We pointed out applications to the weakly screened Yukawa systems, mostly in the context of complex (dusty) plasmas. It is likely that the discussed observation can be also useful for other classical systems with soft repulsive interactions.

\begin{acknowledgements}
SAK present position at the Aix-Marseille-University is supported by the A*MIDEX grant (Nr.~ANR-11-IDEX-0001-02) funded by the French Government ``Investissements d'Avenir'' program. The development of the hybrid approach for the OCP in 3D and 2D was supported by the Russian Science Foundation, Project No. 14-12-01235.
\end{acknowledgements}

\end{document}